\documentclass[12pt]{article}
\usepackage[T1]{fontenc}
\usepackage{tgbonum}
\usepackage{amsmath,slashed,graphicx,xcolor,booktabs,array,
	bbm,url,hyperref}
\usepackage[leftcaption]{sidecap}
\sidecaptionvpos{figure}{c}
\newcommand{\ft}[2]{{\textstyle\frac{#1}{#2}}}

\def \ha{\ft{1}{2}}
\def \dh{\ft{d}{2}}
\def\tr{\mathrm{tr}\,}
\def\cD{\mathcal{D}}
\def\cH{\mathcal{H}}
\def\I{\mathrm{i}}
\def\D{\,\mathrm{d}}
\def\E{\,\mathrm{e}}
\def\Esc{E_\mathrm{sc}}
\def\id{\mathbbm{1}}
\def\wr{w_\mathrm{R}}
\def\mtxt#1{\quad\text{#1}\quad}

\begin{document}
	\begin{titlepage}
		\title {
			\hfill{\small preprint DIAS-STP-85-21}\\[15mm]
			Tunnel Determinants}
		\author{
		     A.W. Wipf\footnote{present address: 
		     	Theoretisch-Physikalisches-Institut, Friedrich-Schiller-Universit\"at Jena, 07743 Jena,
		     	Germany; email:
		     	 wipf@tpi.uni-jena.de; url:
		     	 \url{http://www.tpi.uni-jena.de/qfphysics/homepage/wipf/index.html}}\\[3mm]
			\emph{\small School of Theoretical Physics, Dublin Institute for Advanced Studies,}\\
			\emph{\small 10 Burlington Road, Dublin 4, Ireland}}
		\date{\small received: September 1985}
		\maketitle
\begin{abstract}
			\noindent
Methods for computing the regularized determinants of fluctuation 
operators are being developed. The results follow from the fact that 
these determinants can be expressed by eigenmodes of the fluctuation operator. 
As an application the tunnel determinants of some one- and 
higher-dimensional models are computed. It is shown that every 
fluctuation operator defines a supersymmetric quantum mechanical system.
\end{abstract}
\vskip5mm
\begin{center}\small
		semiclassical expansion  \\[8mm]
	Published version: Nuclear Physics \textbf{B269} (1986) 24-44\\[1mm]
	 doi: 10.1016/0550-3213(86)90363-9\\[1mm]
	  arXiv-ed September 1985 ; \LaTeX-ed January 10, 2022
\end{center}
\end{titlepage}
\tableofcontents

\section{Introduction}\label{sec:introduction}
\begin{figure}[h]
	\centering
	\includegraphics[scale=1]{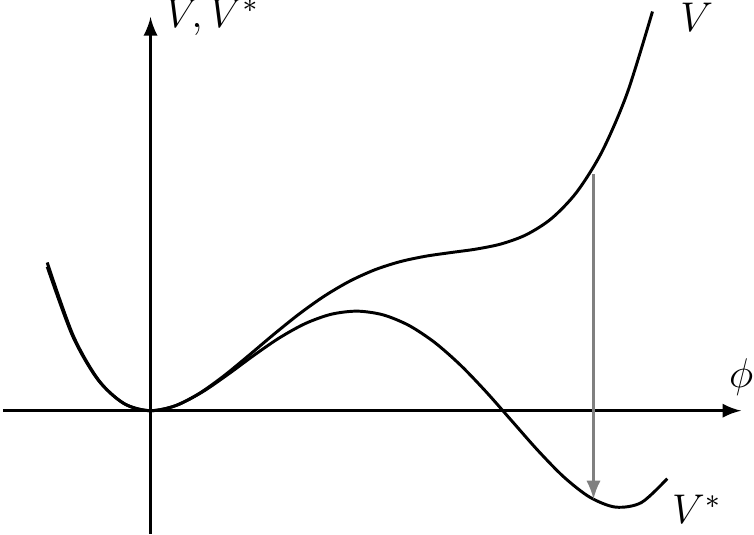}
	\caption{
	A change of the potential which causes a metastable state.
	}\label{fig1}
\end{figure}

The theory of homogeneous nucleation has been a subject of research for 
at least fifty years. An important milestone was reached when Langer \cite{1} 
developed the
field theoretical formalism within the context of statistical mechanics. 
The link with
quantum field theory was provided by Polyakov's paper \cite{2} in which he 
proposed the semiclassical approximation within the euclidean path 
integral formalism. Inspired by Polyakov's work, Callan and 
Coleman \cite{3} constructed the semiclassical
decay amplitude by using the functional integral techniques pioneered by 
Langer. These authors made use of the characterization
\begin{equation}
E_0=-\lim_{T\to\infty}\frac{1}{T}\log\langle x\vert \E^{-TH}\vert x\rangle
=-\lim_{T\to\infty}\frac{1}{T}\bigg(\frac{1}{N}\int_{w(0)=x}^{w(T)=x}
\E^{-S_\mathrm{E}[w]}\cD w\bigg)\label{1.1}
\end{equation}
of the ground state energy of $H$, in order to compute its 
semiclassical approximation $\Esc$. Then they changed the potential 
in $H$; both a typical old potential $V$ and new potential $V^*$ 
are shown in Figure \ref{fig1}. 

As a result $E_0$ ceases to be the ground state 
energy; what is more interesting is that its analytical continuation
develops an imaginary part due to the change of $V$.

In order to compute the analytical continuation of the semiclassical 
approximation $\Esc$, Callan and Coleman integrated over the 
quadratic fluctuations around
multi-bounce configurations. By treating the collective coordinates in a 
suitable manner and assuming that the main contribution comes from 
multi-bounce configurations for which the bounces have negligible 
overlap (dilute gas approximation)
they managed to exponentiate the analytic continuation of the sum
\begin{equation*}
\langle x\vert \E^{-TH}\vert x\rangle_\mathrm{sc}=
\sum_n \langle x\vert e^{-TH}\vert x\rangle_{n\;\mathrm{bounces}}\,,
\end{equation*}
and found an explicit formula for $\Gamma =2 \Im \Esc$.

The analytic continuation of $E_0$ is to be interpreted as the position 
of a second sheet resonance pole of the resolvent of the new 
Hamiltonian $H^*$. Whenever $\Gamma/\vert E_0\vert <1$ one expects the large time 
behaviour
\begin{equation}
p(T)=\big\vert \big(\psi_0,\E^{\I tH^*T}\psi_0\big)\big\vert^2\sim
\E^{-\Gamma\, T}+O\bigg(\frac{\Gamma}{E_0}\bigg)\bigg(\frac{\hbar}{E_0T}\bigg)^\alpha
\label{1.2}
\end{equation}
for the decay of the ground state $\psi_0$ of $H$ under the time evolution 
of $H^*$ \cite{4}. The positive constant $\alpha$ is model dependent. Hence 
we interpret $1/\Gamma$ as the decay time
of the state $\psi_0$.\\[2mm]
At this stage several remarks should be made:
\begin{itemize}
\item Firstly, the decay is exponentially for intermediate times only. For 
$T\to\infty$ an exponential decay is not possible. This is a consequence of Khalfin's theorem \cite{5}
which states that for a Hamiltonian which is bounded below the function 
$(1+x^2)^{-1}\cdot\log p(x)$ is integrable. However, in most 
situations this remark is of minor relevance since we expect that the 
decay already has happened when the second term in \eqref{1.2} dominates the 
first one.

\item Secondly, the Callan-Coleman (CC) formula is valid only 
when the dilute gas condition is fulfilled, i.e. when $\Gamma/V''(0) \ll 1$. 
For the anharmonic oscillator model this inequality excludes the strong 
coupling regime \cite{6}.

\item Thirdly, the derivation given by Callan and Coleman has been criticized. 
For instance, Patrascioiu \cite{7} found a different answer for $\Gamma$ 
for a simple quantum mechanical system by using the complex time method.

\item Finally, it is far from being obvious that the analytic continuation 
of the approximated $E_0$ is a good approximation to the analytic 
continuation of $E_0$.
\end{itemize}
Our interest in the vacuum decay process is due to the crucial role 
it plays in the inflationary cosmological models \cite{8}. There one is 
forced to calculate, or at least estimate, the decay width of a 
metastable state. This symmetric state becomes
metastable, due to the change of the effective potential with 
decreasing temperature. Since the CC approximation for $\Gamma$ 
is the only result known to us which readily can
be generalized to field theory, we adopt a pragmatical attitude and 
assume its validity in what follows. After all, even if the CC 
formula turns out to be not as good as we hope, then the following 
considerations may help to clarify this crucial problem.

The generalization of the CC formula to field theory yields
\begin{equation}
\frac{\Gamma}{V}=\bigg(\frac{S}{2\pi\hbar}\bigg)^{d/2}
\bigg(-{\mathrm{det}}'\frac{-\varDelta+V''(\phi)}{-\varDelta+m^2}\bigg)^{-1/2}
	\exp(-S/\hbar)=\Theta \exp(-S/\hbar)
	\label{1.3}
\end{equation}
for the decay rate per volume and time of the metastable state. 
In this formula $m^2=V''(0)$, where $\phi=0$ is the false vacuum. For 
the computation of the euclidean action
\begin{equation}
S=S[\phi]=\ha\Vert\nabla\phi\Vert^2+\int \D^2 x\,V(\phi)\label{1.4}
\end{equation}
of the solution of the classical field equation (bounce)
\begin{equation}
\varDelta\phi=V'(\phi),\qquad \lim_{\vert x\vert\to\infty}\phi(x)=0
\label{1.5}
\end{equation}
with least action, powerful variational methods are known \cite{9}. In 
cosmological applications one typically uses dimensional arguments 
for a rough estimate of the radiative corrections $\Theta$ to 
$\exp(-S/\hbar)$. It is the purpose of this paper to replace the
rough estimate \cite{10}
\begin{equation}
\Theta\approx \eta M^d\,,\label{1.6}
\end{equation}
where $M$ is a characteristic mass of the theory 
and $\eta$ is a dimensionless number of
order unity, by a more accurate value.

Related results on functional determinants are found in the paper of de 
Vega \cite{11} on the large-N expansion. However, we have to deal with the 
additional difficulty that the fluctuation operator 
$-\varDelta+V''(\phi)= H +m^2$ has $d$ zero modes \cite{3}
which must be dropped in evaluating the tunnel determinant (also
called primed determinant).
This truncation is indicated by the prime in the formula \eqref{1.3}. 
As a by-product we show that every fluctuation operator defines 
a supersymmetric quantum mechanical system. 

In section \ref{sec:tunnel}
we set up  our problem in more detail, give the relevant properties of
functional determinants and relate these objects to scattering 
data of the operator $H$. In two and more dimensions our tunnel determinants 
are UV-divergent. As a regularization we adopt the counterterm method \cite{12}. 
This strategy seems to be more suitable for our task than the Pauli-Vilars, $\zeta$-function or analytic regularizations. Finally we
renormalize the divergent terms in the Feynman-graph expansion of the 
determinant. 

Section \ref{sec:zeromodes} is devoted to the treatment of the zero modes 
of the fluctuation operator. In section \ref{sec:examples} we apply the obtained 
results to some typical models. We find explicit
expressions for one-dimensional and higher-dimensional tunnel 
determinants. The latter correspond to the conformally invariant scalar theories.

In section \ref{sec:supersymmetric} we study the sector with angular momentum 
$j=1$ more carefully and find
that the fluctuation operator defines a supersymmetric quantum mechanical system
on this subspace.

The two appendices are devoted to some more mathematical results. We first
prove that the functional determinant is analytic in $m^2$. Then we show that the
tunnel solution (the bounce) $\phi$ is spherically symmetric, even in cases where 
it is a more-component field.

\section{Tunnel determinants}\label{sec:tunnel}
In defining functional determinants like 
\begin{equation}
w(z)=\det (H-z)(H_0-z)^{-1}\,,\label{2.1}
\end{equation}
where $H_0=-\varDelta$ and $H=H_0+q(x)$ are Schr\"odinger operators, 
one tries to generalize the matrix identity $\log \det B = \tr \log B$ 
to operators. With 
\begin{align}
(H-z)(H_0-z)^{-1}&=1+A(z),\quad \text{where}\nonumber \\
A(z)&=q(x)(H_0-z)^{-1}=q(x)R_0(z)\,,\label{2.2}
\end{align}
one therefore defines
\begin{equation}
\log w(z)=\tr\log \big(1+A(z)\big)\,.\label{2.3}
\end{equation}

For the trace of $\log(1+A(z))$ to exist it suffices that the trace 
of $\vert A(z)\vert$ is finite \cite{12}.
Actually this condition is too strong. For example, the operator
\begin{align*}
B&:\ell_2\mapsto\ell_2\\
&\{x_n\}_1^\infty\mapsto \bigg\{\frac{(-)^{n+1}}{2n-1}x_n\bigg\}_0^\infty
\end{align*}
is not trace class because $\tr\vert B\vert=\sum_1^\infty 1/(2n-1)$ 
is infinite. Nevertheless $\tr\log(1+B)$
exists and is equal $\ft12\log 2$. More generally, the finiteness of 
$\tr B$ implies the finiteness of $\tr \log (1 + B)$ if and only if 
$\tr\vert B\vert^2<\infty$.

The resolvent $R_0(z)$ in \eqref{2.2} is an integral operator with
free two-point Schwinger function $S_2(z;x-y)$ with mass $m^2=-z$ as
integral kernel. Then $A(z)$ is an integral operator as well with 
kernel $A(z;x,y)=q(x)S_2(z;x-y)$, such that
\begin{equation}
\tr A(z)=\int A(z;x,x)\D^dx=\int q(x)\D^dx\, S_2(z;0)\,.\label{2.4}
\end{equation}
In two and more dimensions \cite{13}
\begin{equation*}
S_2(z;\xi)\sim \begin{cases}
c_d\vert\xi\vert^{2-d}&\text{for }\; d>2\,,\\
c\log(\vert\xi\vert)&\text{for }\; d=2\,,
\end{cases}
\end{equation*}
and $\tr A(z)=\infty$ due to the short-distance singularity of the free Schwinger 
function.  Consequently we must regularize the determinant $w(z)$. 
There exist many regularization procedures in field theory which 
easily can be transformed into regularization schemes for
determinants. We will use the counterterm strategy \cite{12} which is 
shortly described as follows.
In the power series expansion\footnote{since $\tr(1+uA)$ is analytic in $u$
we may assume $\vert A(z)\vert <1$.}
\begin{equation}
\log\big( 1+A(z)\big)=-\sum_{n=1}^\infty \frac{(-)^n}{n}\, A^n(z)\,,\label{2.5}
\end{equation}
we denote the sum of the first $p =[\dh]$ terms by $P(A(z))$. 
These are the terms which
have an infinite trace. Now one simply drops this polynomial 
in defining the  \emph{regularized determinant}, i.e.
\begin{equation}
\wr (z) = {\det}_\mathrm{R}(1 + A(z)) = \det \{(1 + A(z)) \E^{-P(A(z))}\}\,.
\label{2.6}
\end{equation}
To define a renormalized determinant we should restore the terms that were
deleted in ${\det}_\mathrm{R}$. It is not legitimate to just leave them out since this would not correspond to local counterterms. For that purpose one should bear in 
mind that one normally
encounters functional determinants as one-loop contributions to effective 
actions. As such they are always accompanied by local one-loop counterterms, 
i.e.
\begin{equation}
\Gamma_\Lambda[\phi] = S[\phi] +\ha\hbar \log {\det}_\Lambda (1 + A(z))+
\hbar S^\mathrm{ct}_\Lambda[\phi]\,.
\label{2.7}
\end{equation}
Here $\Lambda$ indicates a regularization, say a momentum cutoff, 
of the corresponding quantities. Hence one defines the renormalized 
determinant as
\begin{equation}
\log{\det}_\mathrm{ren}\big(1 + A(z)\big) = \lim_{\Lambda\to\infty} 
\{\log {\det}_\Lambda \big(1 + A(z)\big) + 2S^\mathrm{ct}_\Lambda[\phi]\}\,.
\label{2.8}
\end{equation}
To be more specific let us consider 
the model $V(\phi)=\ha m^2\phi^2+\ft{1}{4}\lambda\phi^4$ in $d\leq 4$ dimensions.
In two and three dimensions
\begin{equation}
S^\mathrm{ct}_\Lambda[\phi]=\ha\delta m^2(\Lambda)\int\phi^2
\label{2.9}
\end{equation}
and in four dimensions
\begin{equation}
S^\mathrm{ct}_\Lambda[\phi]=\ha\delta m^2(\Lambda)\int\phi^2
+\ft{1}{4}\delta\lambda(\Lambda)\int\phi^4\,.
\label{2.9prime}
\end{equation}
To proceed we have to choose normalization condition(s). 
It is convenient to use a zero-momentum or constant field renormalization. 
For a constant field $\Phi$ the effective actions $\Gamma$ 
yields the effective potential $U$, i.e. $\Gamma_\Lambda[\Phi]/\text{Vol} 
= U_\Lambda(\Phi)$, and $U_\Lambda$ can be computed 
explicitly in one-loop approximation.

If we take the renormalization conditions
\begin{equation}
\frac{\D^2 U}{\D\Phi^2}\Big\vert_{\Phi=0}=m^2
\label{2.10}
\end{equation}
in two and three dimensions, or
\begin{equation}
\frac{\D^2 U}{\D\Phi^2}\Big\vert_{\Phi=0}=m^2,\qquad
\frac{\D^4 U}{\D\Phi^4}\Big\vert_{\Phi=0}=6\lambda
\label{2.10prime}
\end{equation}
in four dimensions, then
\begin{equation}
{\det}_\mathrm{ren}\big(1+A(z)\big)={\det}_\mathrm{R}\big(1+A(z)\big)
\label{2.11}
\end{equation}
in two and three dimensions and
\begin{equation}
{\det}_\mathrm{ren}\big(1+A(z)\big)={\det}_\mathrm{R}\big(1+A(z)\big)\exp\bigg\{
(3\lambda/4\pi)^2\int\phi^2(x)G(x-y)\phi^2(y)\D^4x\D^4y\bigg\}\,,
\label{2.12}
\end{equation}
in four dimensions,  where
\begin{equation*}
G(\xi)=\frac{1}{(2\pi)^{4}}\int \E^{\I p\xi}\hat G(p)\,\D^4p\,,
\end{equation*}
with
\begin{equation*}
\hat G(p)=\Big(1+\frac{4m^2}{p^2}\Big)^{1/2}\text{Arcoth}\Big(1+\frac{4m^2}{p^2}\Big)^{1/2}-1\,.
\end{equation*}
We see, that if one knows the regularized determinant 
one easily recovers the renormalized one. Therefore, in what follows we are mainly concerned with the problem of calculating the regularized determinants.

In the appendix we show that $\log \wr(z)$ is an analytic function at points which are not in the spectra of both, $H$ and $H_0$. Several authors \cite{14,15} proved that the 
bounce $\phi$ in \eqref{1.5} is spherically symmetric 
for a large class of Higgs potentials. These proofs apply to a 
one-component field. In the appendix we generalize this
result to the case when $\phi$ has several components and the action 
is invariant under rotations of the coordinates and field. 
Therefore we may assume that the function 
$q(x)= V''(\phi)- m^2$ in \eqref{2.2} is spherically symmetric. This allows for an angular momentum expansion of $\wr(z)$. Let
$\cH_j$ denote the subspace with angular momentum $j$. We obtain
\begin{equation}
\wr(z)=\prod_{j=0}^\infty \big\{\det(1+A_{j}(z))\E^{-P(A_{j}(z))}\big\}^{\text{dim}\cH_j}\,,
\label{2.13}
\end{equation}
where
\begin{equation*}
1+A_{j}(z)=(h_j-z)(h^0_j-z)^{-1}\,,
\end{equation*}
and
\begin{equation}
h_j=-\partial_r^2-\frac{d-1}{r}\partial_r+\frac{j(j+d-2)}{r^2}+q(r)
\label{2.14}
\end{equation}
is the radial Schr\"odinger operator on $\cH_j$. Since $h_j$ 
is a one-dimensional operator it follows that $A_{j}(z)$ 
is of trace class. But for trace class operators we have \cite{12}
\begin{equation}
\det\Big\{\big(1+A_{j}(z)\big)\E^{-P(A_{j}(z))}\Big\}
=\det\big(1+A_{j}(z)\big)\E^{-\tr P(A_{j}(z))}
\label{2.15}
\end{equation}
Using the abbreviations
\begin{equation}
w_j(z)=\det\big(1+A_{j}(z)\big)\mtxt{and}p_j(z)=\tr P(A_{j}(z))\,,\label{2.16}
\end{equation}
we finally end up with the expansion
\begin{equation}
\wr(z)=\prod_{j=0}^\infty \big\{w_j(z)\E^{-p_j(z)}\big\}^{\text{dim}\cH_j}\,.
\label{2.17}
\end{equation}
Still we are left with the problem of computing the one-dimensional determinants $w_j(z)$. Here we use the result \cite{11,14} that $w_j(z)$ 
is given by the normalized Jost function of $H$,
\begin{equation}
w_j(z)=F(\mu,k)\mtxt{with} k^2=z,\quad \mu=j-1+\frac{d}{2}\,.\label{2.18}
\end{equation}

The Jost function can be defined in the following way: 
take the solution of the radial Schr\"odinger equation $h_j\psi =k^2\psi$
which fulfills
\begin{equation*}
\lim_{r\to\infty}\E^{\I kr}r^{(d-1)/2}\psi(r)=1\,.
\end{equation*}
Then the unnormalized Jost function is given by
\begin{equation}
f(\mu,k)=2\mu\lim_{r\to 0} r^{j}\psi(r)\,.
\label{2.19}
\end{equation}
To obtain the normalized Jost function one divides $f$ 
by the Jost function $f^0$ of $H_0$, i.e. $F(\mu,k)=f(\mu,k)/f^0(\mu,k)$.

At this stage we may use known properties of Jost functions \cite{16}. For example, $F(\mu,k)$ 
is analytic in $\Im(k)<m$ and $\Re(\mu) > 0$ and its zeros 
on the negative imaginary axis correspond to bound states of $H$. 
Because $w_j(z)$ vanishes exactly if $z$ is a
bound state energy of $H$, this implies that we must choose the 
root $k =\sqrt{z}$ in \eqref{2.18} on the lower half-plane.

\subsubsection*{One dimension} 
In one dimension the product \eqref{2.17} contains only two factors, 
since $\text{dim}\cH_j = 0$ for $j > 1$. No counterterms $p_j$ 
are needed. With $F(\ha,k)F(-\ha,k) = T^{-1}(k)$, where $T$
denotes the transmission coefficient, the expansion \eqref{2.17} reduces to
\begin{equation}
w(z)=\det\bigg(\frac{H-z}{H_0-z}\bigg)=
\frac{1}{T(k)}\,.\label{2.20}
\end{equation}
\subsubsection*{Higher dimensions} 
In higher dimensions we must determine 
the counterterms in the sectors with fixed angular momentum.
For that purpose 
we first expand $F(\mu,k)$ in powers of $q(r)$ \cite{16},
\begin{equation}
F(\mu,k)=\sum_{n=0}^\infty\frac{(-1)^n}{n!}K_n=
\sum \int\frac{(-)^n}{n!}K_n(\mu;r_1,\dots,r_n)q(r_1)\cdots q(r_n)\,,
\label{2.21}
\end{equation}
where
\begin{equation*}
K_n(\mu,r_1,\dots,r_n)=\det
\begin{pmatrix}
C(\mu;r_1,r_1)&\dots&C(\mu;r_1,r_n)\\
\vdots&&\vdots\\
C(\mu;r_n,r_1)&\dots&C(\mu;r_n,r_n)
\end{pmatrix}
\end{equation*}
with 
\begin{equation*}
C(\mu;r,r')=-\ft{\I}{2} \sqrt{rr'}J_\mu(kr_<)H_\mu(kr_>)\mtxt{and}
r_{\stackrel{<}{>}}=\ha(r+r')\pm\ha\vert r-r'\vert\,.
\end{equation*}
According to the described counterterm strategy we next expand
\begin{equation}
\log w_j(z)=\log\bigg(1+\sum_{1}^\infty \frac{(-)^n}{n!}K_n\bigg)
\label{2.22}
\end{equation}
in powers of $q$. The counterterm $p_j$ is equal to the sum of the terms 
up to order $[\dh]$.

Now it is straightforward to express the expansion in terms of the Jost 
function and the $K_n$'s. Especially in three and four dimensions
\begin{equation}
\wr(z)=\begin{cases}
\prod \big\{F(j+\ha,k)\exp(K_1)\big\}^{2j+1}&\text{for }d=3\,,\\
\prod \big\{F(j+1,k)\exp(K_1-\frac{1}{2}[K_2-K_1^2])\big\}^{(j+1)^2}&\text{for }d=4\,.
\end{cases}
\label{2.23}
\end{equation}
One must bear in mind that the $K_n$ in the exponents depend
via $\mu$ on $j$ and $d$. 

\section{Division by the zero modes}\label{sec:zeromodes}
In differentiating the field equation \eqref{1.5} for a spherically symmetric
solution with respect to the radial variable,
\begin{align}
0&=\partial_r\Big(-\partial_r^2\phi-\frac{d-1}{r}\partial_r\phi+V'(\phi)\Big)\nonumber\\
&=-\partial_r^2(\partial_r\phi)-\frac{d-1}{r}\partial_r(\partial_r\phi)+V''(\phi)
\partial_r\phi=(h_1+m^2)\partial_r\phi\,,\label{3.1}
\end{align}
we recognize $\partial_r\phi$ as a zero mode of $h_1+m^2$. 
So there exist $\text{dim}(H_1) = d$ zero modes
of the fluctuation operator with angular momentum $j =1$. 
These are the $d$ Goldstone modes which are due to the translational 
invariance of $S[\phi]$. Therefore $w_1(h)$
vanishes at the interesting point $z=-m^2$. We apply l'Hospital's rule
\begin{equation*}
w_1'(-m^2)=-\lim_{k\to-\I m}\frac{F(\dh,k)}{k^2+m^2}=
\frac{\frac{\D f}{\D k} (\dh,-\I m)}{2\I m f^{0}(\dh,-\I m)}
\end{equation*}
for dividing $w_1$ by this zero mode. Here $f^0$ denotes the free Jost function. 
Next we use the identity \cite{16}
\begin{equation}
\frac{\frac{\D f}{\D k}(\dh,-\I m)}{f(\dh,-\I m)}
=\I\int f(\dh,\I m,r)^2\, r^{d-1}\D r\,,
\label{3.2}
\end{equation}
where $f(\dh,\I m,r)$ is the radial eigenfunction with energy
$-m^2$ which is normalized by
\begin{equation}
\lim_{r\to\infty} r^{(d-1)/2}\E^{mr} f(\dh,\I m,r)=1\,.\label{3.3}
\end{equation}

Because of \eqref{3.1} this eigenstate is proportional to $\partial_r\phi$, i.e.
\begin{equation*}
\sqrt{m} f(\dh,\I m,r)=\frac{1}{B}\frac{\D\phi}{\D r}\,,
\end{equation*}
such that the right-hand side in \eqref{3.2} is equal to a constant times the kinetic energy
of the bounce. Now we use the virial theorem $S[\phi]=(2/d)T[\phi]$, 
which relates the euclidean action and the kinetic energy of $\phi$ \cite{15}, and obtain
\begin{equation}
w_1'(-m^2)=\frac{d}{2m^2B^2 \omega_d}\,\big\vert F(\dh,\I m)\big\vert\,S[\phi],
\mtxt{where}\omega_d=\frac{2\pi^{d/2}}{\Gamma(\dh)}
\label{3.4}
\end{equation}
is the surface area of the $n$-sphere.
So we end up with the angular momentum expansion
\begin{align}
{\det}_\mathrm{R}'\bigg(\frac{-\varDelta+V''(\phi)}{-\varDelta+m^2}\bigg)&=
\big\{w_1'(-m^2)\E^{-p_1(-m^2)}\big\}^d\nonumber\\
&\times \prod_{j\neq 1}\big\{F(j-1+\dh,-\I m)\E^{-p_j(-m^2)} \big\}^{\text{dim}\cH_j}
\label{3.5}
\end{align}
for the regularized tunnel determinant. Besides the Jost function we have to compute,
or at least estimate, the action $S$ of the bounce and the change of normalization between 
$\partial_r\phi$ and $f(\dh,\I m,r)$, i.e.
\begin{equation}
\sqrt{m}B=\lim_{r\to\infty}\Big\{\frac{\D\phi}{\D r} r^{(d-1)/2}\E^{mr}
\Big\}\,.\label{3.6}
\end{equation}

For one-dimensional models we will show how to calculate $B$ explictly. The action
must be computed anyway in \eqref{1.3}. Its value can be estimated with the help of
powerful variational methods \cite{9}.

\section{Examples}\label{sec:examples}
\subsection{Determinants in one dimension}
As explained in section \ref{sec:tunnel} the product \eqref{3.5} contains only two 
factors in one dimension. Since $F(\ha,-\I m)F(-\ha,-\I m)=-2$ we obtain
\begin{equation}
w'(-m^2)=-\frac{2}{(2mB)^2}\,S[\phi]\,.
\label{4.1}
\end{equation}

\begin{figure}[h]
	\centering
	\includegraphics[scale=1]{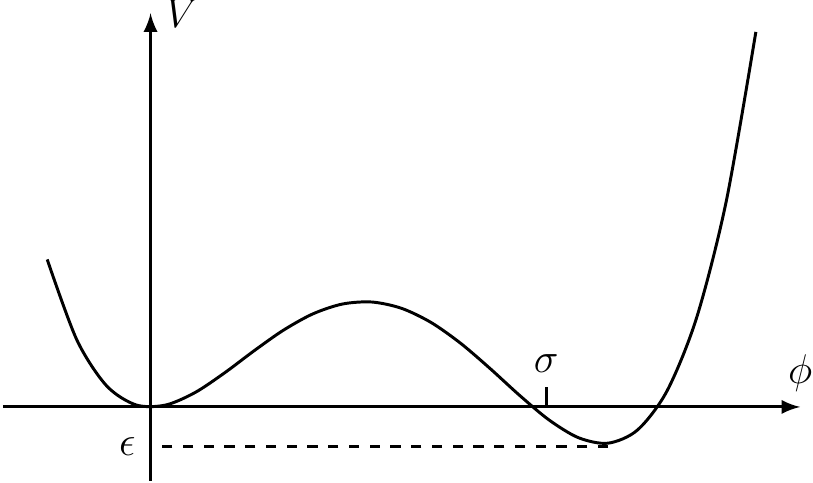}
	\caption{{A typical Higgs potential which gives rise to a fluctuation operator with a negative and at least 
	one zero eigenvalue.}
	}\label{fig2}
\end{figure}

The instanton solution is the inverse function of
\begin{equation}
x(\phi)=\int_0^\phi \big\{2V(\phi') \big\}^{-1/2}\D\phi'\label{4.2}
\end{equation}
and the action of the bounce is
\begin{equation}
S[\phi]=2\int_0^\sigma \big\{2V(\phi') \big\}^{1/2}\D\phi'\,.\label{4.3}
\end{equation}
There is a factor $2$ since a bounce consists of an instanton and anti-instanton.
The constant $B$ in \eqref{3.6} becomes
\begin{equation}
B=\lim_{\phi\to 0}\sqrt{m}\,\phi(x)\E^{mx(\phi)}\,.
\label{4.4}
\end{equation}
The point $\sigma$ up to which we integrate in \eqref{4.3} is the value to 
which the particle tunnels. A typical potential is shown in Figure~\ref{fig2}.
Let us illustrate these with three examples.

\subsubsection*{Unstable potentials}

\begin{figure}[h!]
	\centering
	\includegraphics[scale=0.95]{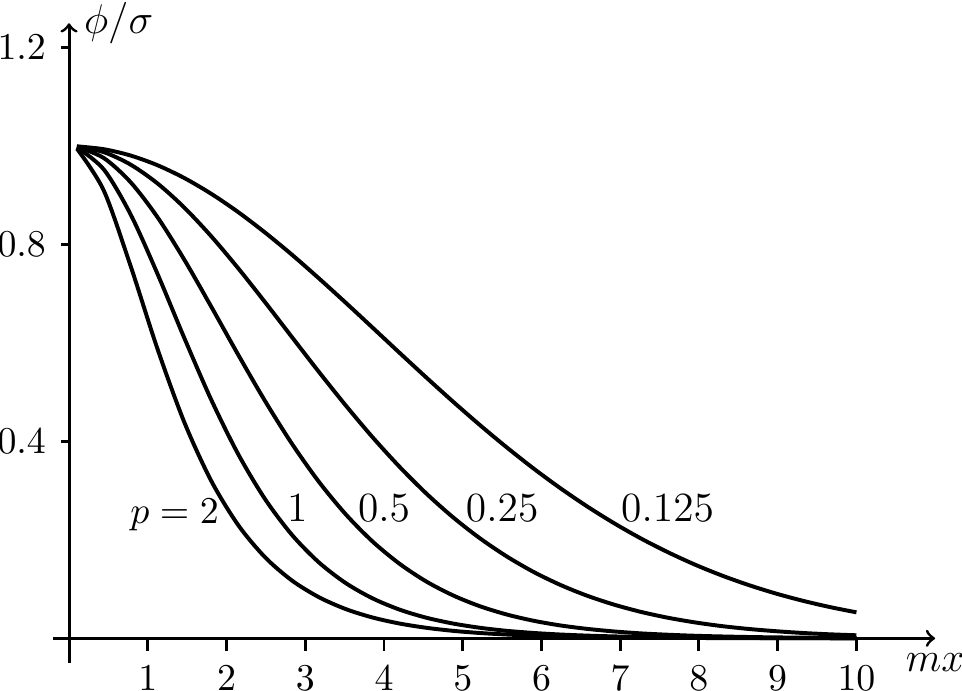}
	\caption{
		The tunnel solution for the model (\ref{model1}) for different $p$'s.
	}\label{fig3}
\end{figure}

\begin{figure}[h!]
	\centering
		\includegraphics[scale=0.95]{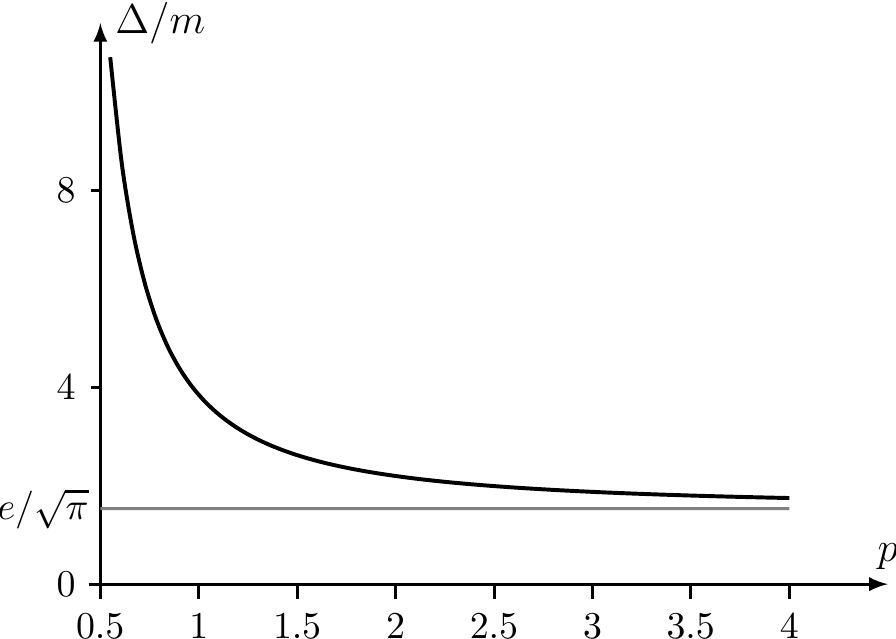}
	\caption{
		 The ratio $\Delta=\Theta/\E^{-S}$ for the model (\ref{model1}) as function of $p$ for ($m\sigma^2=1$).
	}\label{fig4}
\end{figure}

As model for unstable potentials we use
\begin{equation}
V(\phi)=\frac{m^2}{2}\phi^2\Big(1-\frac{\phi^p}{\sigma^p}\Big),\quad p>0\,.\label{model1}
\end{equation}
It admits the tunnel solution $\phi(x)=\sigma\cosh^{-2/p}(\ft{pm}{2}x)$ 
with action 
\begin{equation*}
S=\frac{2m\sigma^2}{p+4}\, B\big(\ha,\ft{2}{p}\big)\,,
\end{equation*}
where $B(a,b)$ denotes the beta-function.
For the constant in \eqref{4.4} we find $B=2^{2/p}\sigma\sqrt{m}$ such that the formula \eqref{4.1} gives the tunnel determinant
\begin{equation}
{\det}'\bigg(\frac{-\partial_x^2+m^2-\ha (p+1)(p+2)m^2\cosh^{-2}(\frac{pm}{2} x)}{-\partial_x^2+m^2}
\bigg)
=-\frac{1}{m^2}\frac{B\big(\ha,\ft{2}{p}\big)}{(p+4)\,2^{4/p}}\,.
\label{4.5}
\end{equation}
Inserting into \eqref{1.3}
we obtain the quantum correction to the decay rate
\begin{equation}
\Theta=
\bigg(\frac{S}{2\pi\hbar}\bigg)^{1/2}
{\mathrm{det}}'\bigg(\frac{-\partial_x^2+V''(\phi)}
{-\partial_x^2+m^2}\bigg)^{-1/2}=m\sigma \Big(\frac{m}{\pi}\Big)^{1/2}\,2^{2/p}\,.
\end{equation}
Now let $p$ approach $0$. Then the width of the bounce 
gets larger, for example
$\phi(x)=\sigma/e$ for $x=(1+\sqrt{4/p})/m$. The limit $p\to 0$
is somehow similar to the thin wall limit in higher dimensional theories \cite{3}. 
There the width of the bounce diverges in the limit of a vanishing 
potential energy density difference $\epsilon$ between the true and false vacuum. 
In using the asymptotic expansion for the beta-function
$B(\ha,\ft{2}{p})$ we find for the classical factor in the decay rate \eqref{1.3}
\begin{equation*}
\E^{-S}\sim \exp\bigg\{-\Big(\frac{p\pi}{2}\Big)^{1/2}\,\frac{2m\sigma^2}{p+4}
\bigg\}\,.
\end{equation*}
In Figures \ref{fig3} and \ref{fig4} we show the bounces for different 
values of $p$ and the ratio $\Theta/\exp(-S)=\Delta$ as function of $p$, respectively. 
We recognize that in the "thin-wall region" $p\to 0$ the quantum corrections dominate the classical contribution. We conclude that the semiclassical expansion 
breaks down for small values of $p$.


\subsubsection*{Double well potential}
We consider the quartic double well potential
\begin{equation}
V(\phi)=\frac{m^2\phi^2}{2}\Big(1-\frac{\phi}{\sigma}\Big)^2\,,\label{model2}
\end{equation}
which admits the bounce $\phi(x)=\ha\sigma(1+\tanh\ha mx)$. Although the Goldstone 
mode has angular momentum $j=0$, the formula \eqref{4.1} still holds (up to a sign). 
With $S=\ft{1}{6}m\sigma^2$ and $B =\sqrt{m}\sigma$ we find
\begin{equation}
{\det}'\bigg(\frac{-\partial_x^2+m^2-\ft{3}{2}m^2\cosh^{-2}(\ha mx)}{-\partial_x^2+m^2}\bigg)
=\frac{1}{24m^2}\,.
\label{4.6}
\end{equation}
\subsubsection*{Sine-Gordon potential}
The periodic potential
\begin{equation}
V(\phi)=\bigg(\frac{m\sigma}{2\pi}\bigg)^2\bigg(1-\cos\frac{2\pi\phi}{\sigma} \bigg)
\label{model3}
\end{equation}
has the instanton solution $\phi(x)=(2\sigma/\pi)\tan^{-1}\{\exp(mx)\}$
with associated bounce
action $S=(2\sigma/\pi)\sqrt{m}$ and constant $B=2m\sigma^2/\pi^2$. 
Hence we obtain for the tunnel determinant
\begin{equation}
{\det}'\bigg(\frac{-\partial_x^2+m^2-2m^2\cosh^{-2}(mx)}{-\partial_x^2+m^2}\bigg)
=\frac{1}{4m^2}\,.
\label{4.7}
\end{equation}
\subsubsection*{Fluctuation operators with Eckardt potentials}
\begin{figure}[h]
	\centering
	\includegraphics[scale=1]{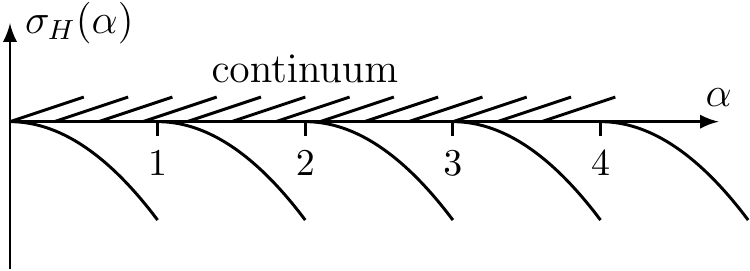}
	\caption{The spectral flow with $\alpha$ for the Hamiltonian 
		with Eckardt potential \eqref{4.8}}
	\label{fig5}
\end{figure}

Actually, the tunnel determinants 
(\ref{4.5},\ref{4.6},\ref{4.7}) are obtained more easily since for the Eckardt potential
\begin{equation}
q(x)=-\frac{\alpha(\alpha+1)}{\rho^2}\cosh^{-2}\Big(\frac{x}{\rho}\Big)\label{4.8}
\end{equation}
the transmission coefficient in \eqref{2.20} is explicitly known \cite{17}
\begin{align}
T^{-1}(\I k,\alpha)&=
\det\bigg(\frac{-\partial_x^2-k^2-\alpha(\alpha+1)/\rho^2\cdot\cosh^{-2}(x/\rho)}{-\partial_x^2-k^2}
\bigg)\nonumber\\
&=\frac{\Gamma(\I k\rho+1)\,\Gamma(\I k\rho)}{\Gamma(\I k\rho+1+\alpha)\,
	\Gamma(\I k\rho-\alpha)}\,.
\label{4.9}
\end{align}
It gives rise to a spectral flow with respect to the parameter $\alpha$ 
as shown in Figure \ref{fig5}. 

For $\alpha =1+2/p$ and $m\rho=2/p$ is the potential (\ref{4.8}) 
just the potential in the fluctuation operator \eqref{4.5}, 
for $\alpha=2$ and $m\rho=2$  the potential in the fluctuation operater 
\eqref{4.6} and for $\alpha=1$ and $m\rho= 1$ the potential in the operator \eqref{4.7}. 
Now one can directly divide \eqref{4.9} by the eigenvalue which vanishes for $k=-\I m$
and perform the limit $k\to -\I m$ for the quotient to recover the
results (\ref{4.5},\ref{4.6},\ref{4.7}) for the tunnel determinants.
\subsection{Determinants in higher dimensions} 
Although the conformally invariant model in $d$ dimensions
\begin{equation}
V(\phi)=-\frac{\lambda}{d_c}\phi^{d_c}\,,\mtxt{with} d_c=\frac{2d}{d-2}\label{4.10}
\end{equation}
shows no tunnel effect in the usual sense, the corresponding fluctuation operator
defines a tunnel determinant. Due to the scale invariance of this theory the instanton
solution \cite{18}
\begin{equation}
\phi(x)=\bigg\{\frac{2d}{\lambda d_c} \bigg\}^{d/d_c}\bigg\{\frac{\alpha}{1+\alpha^2
(x-x_0)^2}\bigg\}^{d/d_c}
\label{4.11}
\end{equation}
depends on the additional parameter $\alpha$. 
Thus the fluctuation operator $(\alpha=1,x_0=0)$
\begin{equation}
H=-\varDelta-\frac{d(d+2)}{(1+r^2)^2}\label{4.12}
\end{equation}
has the zero modes
\begin{align}
\partial_\alpha\phi&=c\frac{1-r^2}{1+r^2}\,\phi\approx r^{2-d}\mtxt{for}r\to\infty\,,
\nonumber\\
\partial_r\phi&=\tilde c\frac{r}{1+r^2}\,\phi\approx r^{1-d}\mtxt{for}r\to\infty\,,
\label{4.13}
\end{align}
in the sectors with $j=0$ and $j=1$, respectively. Because the Jost functions 
are only known at $k=0$ \cite{18}
\begin{equation}
F(\mu,k=0)=\frac{\Gamma(\mu)\Gamma(\mu+1)}{\Gamma(\mu-\dh)\Gamma(\mu+\dh+1)}\,,
\label{4.14}
\end{equation}
we use a slightly different method for dividing by the zero modes.

Let $E(\mu)$ be an eigenvalue of $h_j$, where $\mu$ and $j$ are
related as in \eqref{2.18}. Then the determinant $w_j(z)=F(\mu,k)$  of the fluctuation
operator on $\cH_j$ vanishes. It follows 
that $F(\mu,k) =\{E(\mu)^{1/2}-k\}\tilde F(\mu,k)$. 
As a consequence we find
\begin{equation}
-E(\mu)^{-1/2}\,\frac{\partial F}{\partial k}\Big\vert_{(\mu,E(\mu)^{1/2})}
=\frac{1}{E'(\mu)}\,\frac{\partial F}{\partial\mu}\Big\vert_{(\mu,E(\mu)^{1/2})}
\,.\label{4.15}
\end{equation}
In our case $E(\mu)=0$ and since $F(\mu,0)$ is explicitly given 
in \eqref{4.14} it  suffices to calculate $E'(\mu)$ for computing the primed determinant
\begin{equation*}
E(\mu)^{-1/2}\,\frac{\partial F}{\partial k}\Big\vert_{(\mu,E(\mu)^{1/2})}\,.
\end{equation*}
It is enough to calculate $E(\mu)$ in first order in the perturbation 
$h_{j'}-h_j=\mu/r^2$. The unperturbed wave functions are the zero modes \eqref{4.1}. 
So we find for $j=0$ or equivalently $\mu=\dh-1$ and for  $d>4$
\begin{equation}
E'(\dh-1)=E_0'=\Big\langle\frac{1}{r}\Big\rangle_{\partial_\alpha\phi}
=\frac{d-4}{d+2}\,,
\label{4.16}
\end{equation}
and for $j=1$ or equivalently $\mu=\dh$ and $ d>2$
\begin{equation}
E'(\dh)=
E_1'=\Big\langle\frac{1}{r}\Big\rangle_{\partial_r\phi}
=\frac{d-2}{d}\,.
\label{4.17}
\end{equation}
The IR-divergence of the theory shows itself in the non-normalizability of the
Goldstone mode $\partial_\alpha\phi$ or equivalently in the divergence of 
$E_0'^{-1}$ in less than four dimensions. It should be canceled by the 
infrared divergent mass counterterm
\begin{equation*}
\ha\delta m^2\int \phi^2=c\delta m^2\;\frac{d-2}{d-4}\,.
\end{equation*}
With (\ref{4.14},\ref{4.15}) we end up with
\begin{equation*}
	w_0'=-\frac{1}{E_0'}\,\frac{B\big(\dh,\dh\big)}{(d-2)^2}\mtxt{and}
	w_1'=\frac{1}{E_1'}\,B\big(\dh,\dh\big)\,.
\end{equation*}
The counterterms $p_j(0)$ in \eqref{3.5} are easily found by expanding 
$\log F(\mu)$ in powers of the potential $q(r)$. For the fluctuation operator 
\eqref{4.12} we therefore define $\alpha=\alpha(\xi)$
through $\xi d(d +2)=\alpha(\alpha+2)$ and expand $\log F_\alpha(\mu))=C-\log 
B(\mu-\ha\alpha,\mu+1+\ha\alpha)$
in powers of $\xi$. Clearly the sum of the first $[\dh]$ terms at 
$\xi=1$ is equal to $p_j(0)$. 

We find in three dimensions
\begin{equation*}
\wr'(0)=-\frac{1}{E_0'E_1'^3}\frac{\pi}{4} \E^{15}\Big(\frac{\pi}{48}\Big)^3
\prod_{2}^\infty \bigg\{ \frac{B(j+\ha,j+\ft{3}{2})}{B(j-1,j+3)}\,
\exp\Big(\frac{15}{4j+2}\Big)\bigg\}^{2j+1}
\end{equation*}
and in four dimensions
\begin{equation*}
\wr'(0)=-\frac{1}{E_0'E_1'^4}\frac{\E^{198}}{12}\Big(\frac{1}{48}\Big)^4
\prod_{2}^\infty \bigg\{ \frac{j(j-1)}{(j+2)(j+3)}
\exp\Big(\frac{6}{j+1}+18\Big)\bigg\}^{(j+1)^2}\,.
\end{equation*}

\section{The supersymmetric sector}\label{sec:supersymmetric}
\begin{figure}[h]
	\centering
	\includegraphics[scale=1]{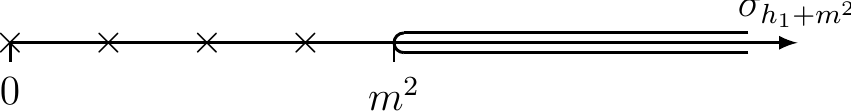}
	\caption{The spectrum of the supersymmetric Hamiltonian
		$h_1+m^2$}
	\label{fig6}
\end{figure}

We have seen that
\begin{equation}
h_1+m^2=-\partial_r^2-\frac{d-1}{r}\partial_r+\frac{d-1}{r^2}+V''(\phi(r))
\label{5.1}
\end{equation}
has always a zero mode - the Goldstone mode $\partial_r\phi$. 
The $L_2$ norm of this mode is proportional to the action of the instanton
and hence is finite. Furthermore, since the Goldstone mode has 
no zeros in $(0,\infty)$ it is the normalizable ground state
of $h_1$. For positive energies the spectrum of $h_1$ is continuous. 
So the spectrum of $h_1 + m^2$ looks as shown in Figure \ref{fig6}.

After these remarks it is not very surprising that $h_1+m^2$ is always a component
of a supersymmetric Hamiltonian \cite{19}, i.e. $h_1+m^2=S^\dagger S$ , 
where $S=-r^{-D}\partial_r r^D+W,\;
D=\ha(d-1)$ and $W=(\D/\!\D r) \log(r^D\partial_r\phi)$.

Things become more transparent if we define the unitary map
\begin{align}
U:\quad &L_2(R^+,r^{d-1}\D r)\longmapsto L_2(R,\E^{2x}\D x)\,\label{5.2}\\
 &\psi(x)\longmapsto\exp\Big(\frac{d}{d_c}x\Big)\psi\big(\E^x\big)\,,\nonumber
\end{align}
so that the transformed Hamiltonians read
\begin{equation}
Uh_j U^{-1}=\E^{-2x}\Big\{-\partial_x^2+\mu^2
+\E^{2x}\Big(V''\big(h(x)\big)-m^2\Big) \Big\}\,,\label{5.3}
\end{equation}
with  $h(x)=\phi(\!\E^x)$ and $\mu=j-1+\dh$ was introduced in \eqref{2.18}.

In using the field equation for the instanton solution it is easy to cheek the identity
\begin{equation}
\E^{2x}V''(h(x))=w^2(x)+w'(x)-\big(\ft{d}{2}\big)^2\,,\label{5.4}
\end{equation}
where
\begin{equation}
w(x)=\partial_x\log \partial_x h+\dh-2\,.\label{5.5}
\end{equation}
Thus we end up with
\begin{equation}
U(h_j+m^2)U^{-1}=\E^{-2x}\big\{S^\dagger S+\mu^2-\big(\ft{d}{2}\big)^2\big\}=\E^{-2x}k_j\,,
\label{5.6}
\end{equation}
where $S=-\partial_x+w(x)$ and $S^\dagger=\partial_x+w(x)$ is its adjoint
in $L_2(R,\D x)$.

Especially interesting is the case $j=1$ or equivalently $\mu=\dh$
because $k_1=S^\dagger S$ 
is part of the supersymmetric Hamiltonian
\begin{equation}
K=\ha(Q_1^2+Q_2^2)=(-\partial_x^2+w^2)\id+w'\sigma_3
=\begin{pmatrix}S^\dagger S&0\\ 0&SS^\dagger\end{pmatrix}
\,,\label{5.7}
\end{equation}
where $Q_1=\sigma_1p+\sigma_2 w$ and $Q_2=\sigma_2 p-\sigma_1 w$.

Because the spectra of $S^\dagger S$ and $SS^\dagger$ are,
up to possible zero modes, exactly the same we naively expect
\begin{equation}
{\det}'\bigg(\frac{S^\dagger S}{SS^\dagger}\bigg)=
{\det}'\bigg(\frac{-\partial_x^2+w^2+w'}{-\partial_x^2+w^2-w'}\bigg)
\stackrel{?}{=}1\,,
\label{5.8}
\end{equation}
or, because
\begin{align*}
w_j(-m^2)&=\det(h_j+m^2)(h_j^0+m^2)^{-1}\\
&=\det\big(U^{-1}\E^{-2x}k_j UU^{-1}k_j^{0-1} \E^{2x} U\big)=
\det k_j k_j^{0-1}
\end{align*}
that
\begin{equation}
w_1'(-m^2)={\det}'\bigg(\frac{S^\dagger S}{S_0^\dagger S_0}\bigg)\stackrel{?}{=}
\det\bigg(\frac{SS^\dagger}{S_0^\dagger S_0}\bigg)\,.
\label{5.9}
\end{equation}
We will see that \eqref{5.8} generally is not true, due to the continuous part 
in the spectrum of $h_1$. Or in other words, the Hamiltonian \eqref{5.7} defines an 
unpairing supersymmetric quantum mechanics.

To be more specific we apply these results to the conformally invariant model
\eqref{4.10}. With (\ref{5.5},\ref{5.6}) and \eqref{4.11} we obtain for the transformed radial 
Hamiltonians
\begin{align*}
k_j&=-\partial_x^2+\mu^2-\ft{1}{4}d(d+2)\cosh^{-2}(x)\,,\\
k_j^0&=-\partial_x^2+\mu^2
\end{align*}
and hence with \eqref{2.20}
\begin{equation}
\det\bigg(\frac{S^\dagger S-z}{SS^\dagger-z}\bigg)
=\det\bigg(\frac{S^\dagger S-z}{S_0^\dagger S_0-z}\bigg)\,
\det\bigg(\frac{S_0^\dagger S_0-z}{SS^\dagger-z}\bigg)=
\det\bigg(\frac{T_{SS^\dagger}(k)}{T_{S^\dagger S}(k)}\bigg)\,.\label{5.10}
\end{equation}
For computing the ratio of the two transmission coefficients we assume that 
$\psi$ is a solution of $S^\dagger S\psi=k^2\psi$  with boundary conditions
\begin{equation*}
\psi\to \begin{cases}\E^{\I kx}+R_{S^\dagger S}(k)\,\E^{-\I kx}&\text{for }
x\to+\infty\,,\\
T_{S^\dagger S}(k)\,\E^{\I kx}&\text{for }x\to-\infty\,.
\end{cases}
\end{equation*}
Then $S(S^\dagger S)\psi=(SS^\dagger)S\psi=k^2 S\psi$, what means that $S\psi$ is an
eigenfunction of $SS^\dagger$ with the same energy $k^2$. Since
\begin{equation*}
S\psi\to \begin{cases}(-\I k+w(\infty))\E^{\I kx}+(\I k+w(\infty))R_{S^\dagger S}(k)\,\E^{-\I kx}&\text{for }
x\to +\infty\\
(-\I k+w(-\infty))T_{S^\dagger S}(k)\,\E^{\I kx}&\text{for }x\to-\infty\,,
\end{cases}
\end{equation*}
we obtain
\begin{equation}
\det \bigg(\frac{S^\dagger S-z}{SS^\dagger-z}\bigg)
=\frac{\I k-w(-\infty)}{\I k-w(+\infty)}\,.
\label{5.11}
\end{equation}
Again we use l'Hospital's rule to compute the primed determinant \eqref{5.8}. With
$w =w(\infty=-w(-\infty)$ we end up with
\begin{equation}
{\det}'\bigg(\frac{S^\dagger S}{SS^\dagger}\bigg)=\frac{1}{4 w^2}\label{5.12}
\end{equation}
instead of \eqref{5.8}, or with
\begin{equation}
w_1'(-m^2)=\frac{1}{4w^2}\det\bigg(\frac{SS^\dagger}{S_0S^\dagger_0}\bigg)\label{5.13}
\end{equation}
instead of \eqref{5.9}.

Phrased in other words, the Witten index $\text{dim}\,\text{ker}(S)-\text{dim}\, 
\text{ker}(S^\dagger)$ is not given by the trace $\tr\{\sigma_3\exp(-tK)\}$.
This can be seen from
\begin{align*}
-\int_0^\infty \frac{\D t}{t}\Big\{\tr\sigma_3 \E^{-t K}-\text{index}(S)\Big\}
&=-\int_0^\infty \frac{\D t}{t}\big\{\tr\big(\E^{-t S^\dagger S}-\E^{-t SS^\dagger}\big)-\text{index}(S)\big\}\\
&=\log\,{\det}'\,\frac{S^\dagger S}{SS^\dagger}=-\log (4w^2)\neq 0\,,
\end{align*}
and is due to the continuous part of the spectrum of $S^\dagger S$ 
or the fact that the theory is defined on an open space. For the 
model \eqref{4.12} we have $w(x)=\dh \tanh x$ and the corresponding identity
\begin{equation*}
{\det}'\bigg(\frac{-\partial_x^2+\ft{1}{4}d^2-\ft{1}{4}d(d+2)\cosh^{-2}x}
{-\partial_x^2+\ft{1}{4}d^2-\ft{1}{4}d(d-2)\cosh^{-2}x}\bigg)=\frac{1}{d^2}
\end{equation*}
can be checked directly by using \eqref{4.9}.
\section{Summary}\label{sec:summary}
In this paper I have presented a method for calculating the tunnel determinant
of fluctuation operators. The result for one-dimensional systems, eq. \eqref{4.1}, has been
applied to a one-parametric class of unstable potentials, the anharmonic double
well potential and finally the sine-Gordon potential. We found that, in cases where
the width of the tunnel solution becomes large, the one-loop quantum corrections
dominate the classical result.

Next we applied the result \eqref{3.5} for the regularized determinant to the fluctuation
operator of the conformally invariant model \eqref{4.10}. We computed $w'_\mathrm{R}$ in any dimension
and gave the explicit expressions in three and four dimensions.

Our methods are applicable to any model for which we know, or at least can
estimate, the Jost function. In \cite{14} the Jost function for the "thin-wall" model
\begin{equation}
q(r)=\begin{cases}\text{const}&\text{for }r<R\,,\\
0&\text{for }r>R\,,
\end{cases}
\label{6.1}
\end{equation}
was used for computing the tunnel determinant of the corresponding fluctuation
operator. Again the quantum corrections compete with the classical part for large
values of $R$.

For both models \eqref{4.5} and \eqref{6.1} we found an exponential dependence of
the quantum corrections on the width of the tunnel-solution, i.e.
\begin{equation*}
\Theta\approx \text{const }\exp\big\{(m\text{ width}^{-d_c/d})^\alpha \big\}\,,
\end{equation*}
where $\alpha$ depends on the dimension of spacetime. Thus the rough dimensional
estimate \eqref{1.6} gives a wrong result in this interesting limit.

With \eqref{5.13} we gave an alternative method for computing the primed determinant
in the $j=1$ sector. The primed determinant of $h_1$ is, up to a surface term, 
given by the unprimed determinant of its "supersymmetric partner". As a by-product we
showed that every fluctuation operator defines an unpairing supersymmetric 
quantum mechanical system.

The author wishes to thank J. Alberty, L. O'Raifeartaigh, N. Straumann and 
H. Yoneyama for useful discussions.

\appendix
\section{Tunnel determinants are analytic}
In this appendix we show that
\begin{equation*}
\log \wr(z)=\tr\big\{ \log\big(1+A(z)\big)-P(A(z))\big\}\,,
\end{equation*}
where $A(z)=q(x)R_0(z)$ and
\begin{equation*}
P(A(z))=A(z)-\ha A^2(z)+\ft{1}{3}A^3(z)-\dots -\ft{(-)^p}{p}A^{p}(z),\quad
p=[\dh]\,,
\end{equation*}
is an analytic function of $z$ at points which are not in the spectra of both, $H$ 
and $H_0$. In the following we shall often skip the argument $z$
in order to simplify the notation.
We first observe that because of the identities
\begin{align*}
\frac{\D A}{\D z}&=AR_0\,,\\
R_0(1+qR_0)^{-1}qR_0&=R_0\big\{(H-z)R_0\big\}^{-1}A=R_0R_0^{-1}RA=RA,\quad
R=\frac{1}{H-z}\,,
\end{align*}
we have
\begin{equation*}
\frac{\D}{\D z}\log \wr=(-)^{p+2}\tr R_0(1+A)^{-1}A^{p+1}=(-1)^p\tr(RA^{p+1})\,.
\end{equation*}
Now we use the analytic nature of the resolvent operator, i.e.
\begin{equation*}
R(z)=\sum_0^\infty (z-z_0)^n R(z_0)^{n+1}
\end{equation*}
to show that $(\!\D/\!\D z)\log\wr$ has an absolute convergent Taylor expansion. 
In
\begin{align*}
\tr R(z)A^\alpha&=\sum_{m,n_1,\dots,n_\alpha}(z-z_0)^{m+n_1+\dots+n_\alpha}
\,\tr R(z_0)^{m+1}qR_0(z_0)^{n_1+1}\cdots qR_0(z_0)^{n_\alpha+1}\\
&=\sum_k a_k (z-z_0)^k\,,
\end{align*}
where $\alpha=p+1$, and the coefficients $\alpha_k$ are bounded by
\begin{align*}
\vert a_k\vert&\leq \sum_{m+n_1+\dots+n_\alpha=k}
\big\Vert R(z_0)^{m+1}qR_0(z_0)^{n_1+1}\cdots qR_0(z_0)^{n_\alpha+1}\big\Vert_1\\
&\leq \sum_{m+n_1+\dots+n_\alpha=k}
\big\Vert R(z_0)\big\Vert^{m+1}\,\big\Vert qR_0(z_0)^{n_1+1}\cdots qR_0(z_0)^{n_\alpha+1}\big\Vert_1\,.
\end{align*}

We used the inequality $\Vert A_1A_2\Vert_1 \leq \Vert A_1\Vert\,\Vert A_2\Vert_1$
for a bounded $A_1$ and a trace class operator $A_2$. Here $\Vert \cdot \Vert$ 
denotes the usual operator norm and 
$\Vert A\Vert^p_p =\tr(AA^\dagger)^{p/2}$.
Next we iterate the inequality (see, for example \cite{12})
\begin{equation*}
\Vert A_1A_2\Vert_q\leq \Vert A_1\Vert_{q_1}\,\Vert A_2\Vert_{q_2},\mtxt{where}
\frac{1}{q}=\frac{1}{q_1}+\frac{1}{q_2},\quad A_k\in P_{q_k}\,,
\end{equation*}
and obtain
\begin{equation*}
\Vert A_1\cdots A_\alpha\Vert_q\leq \Vert A_1\Vert_{q_1}\Vert A_2\Vert_{q_2}\cdots\Vert A_\alpha\Vert_{q_\alpha},\mtxt{with}
\frac{1}{q}=\frac{1}{q_1}+\frac{1}{q_2}+\dots+\frac{1}{q_\alpha}\,,
\end{equation*}
Now we choose $q=1$ and $q_i=\alpha$ which yields
\begin{align*}
\vert a_k\vert&\leq \sum \Vert R\Vert^{m+1}\,\Vert qR_0^{n_1+1}\Vert_\alpha\cdots 
	\Vert qR_0^{n_\alpha+1}\Vert_\alpha\\
		&\leq \sum \Vert R\Vert^{m+1}\,\Vert R_0\Vert^{n_1+\dots +n_\alpha}\cdot 
			\Vert qR_0\Vert_\alpha\;.
\end{align*}

With the definition $C:=\max\{\Vert R(z_0)\Vert,\Vert R_0(z_0)\Vert\}$,
\begin{equation*}
\vert a_k\vert\leq C^{k+1}\Vert A(z_0)\Vert^\alpha_\alpha
\sum_{m+n_1+\dots+n_\alpha=k}1=
C^{k+1}\Vert A(z_0)\Vert_\alpha^\alpha \frac{(\alpha+1)\cdots (\alpha+k)}{k!}\,.
\end{equation*}
On the other side
\begin{equation*}
\Vert A(z_0)\Vert^\alpha_\alpha\leq (2\pi)^{-d/\alpha}\Vert q\Vert_\alpha^\alpha\,
\Vert \tilde S_2\Vert^\alpha_\alpha\,,
\end{equation*}
where $\tilde S_2(p)=(p^2-z_0)^{-1}$ is the free propagator in momentum space. 
The right-hand side is finite for $\alpha>\dh$, $z_0\in\rho(H_0)$ 
and sufficiently fast decaying $q$. The radius of convergence $R=1/C$
is finite for $z_0\in \rho(H)\cap \rho(H_0)$. This finishes the proof
of our statement

\section{Multifield bounces are spherically symmetric}
The results in this paper heavily rely on the spherical symmetry of the bounce.
Several authors \cite{14,15} showed that a one-component tunnel solution is
indeed spherically symmetric. In the present appendix we generalize this result 
to the case when the action is invariant with respect to some
non-trivial group, i.e. with respect to unitary transformations of the
form $\phi\mapsto U(g)\phi$. We assume that the internal symmetry
group act transitively on the sphere defined by fields of fixed
length $\Vert\phi\Vert$. Hence for every 
$x\in R^d$ we can find a transformation $g(x)$ such that the transformed field
\begin{equation}
U(g(x))\,\phi(x)=\Vert\phi(x)\Vert e_1=\chi(x)\label{B.1}
\end{equation}
points always into a fixed direction $e_1$ in field space.

By assumption the potential energy is invariant such that
\begin{equation}
V[\phi]=V[\chi]\,.
\label{B.2}
\end{equation}
The kinetic energy is not invariant, since $g(x)$ depends on the
coordinates. But we can show that the kinetic energy of $\chi$ is 
less or equal than the kinetic energy of $\phi$. For that purpose we use
\begin{equation*}
\nabla\chi=\nabla(\phi,\phi)^{1/2}e_1=(\phi,\phi)^{-1/2}\,\Re(\phi,\nabla\phi)\,e_1\,,
\end{equation*}
and find
\begin{align*}
(\nabla\chi,\nabla\chi)&=(\phi,\phi)^{-1}\big\{\Re(\phi,\nabla\phi)\big\}^2
\leq (\phi,\phi)^{-1}\vert (\phi,\nabla\phi)\vert^2\\
&\leq (\phi,\phi)^{-1}(\phi,\phi)(\nabla\phi,\nabla\phi)=(\nabla\phi,\nabla\phi)\,,
\end{align*}
where we only used the Schwartz inequality. After integrating over $R^d$ we end up
with
\begin{equation}
T[\chi]\leq T[\phi]\,.\label{B.3}
\end{equation}
We use the symbol $\cH$ for the space of all multi-component Higgs fields  
(typically a  Sovolev space) and 
let $\cH_1$ denote the subspace of fields which point into the $e_1$ direction.

We will use the fact that an extremum of $S$, restricted to $\cH_1$, 
is also an extremum of $S$ on the whole space $\cH$. This is an immediate 
consequence of the principle of symmetric criticality \cite{20}. 
This theorem states that if $S$ is invariant under the action
of a compact group $G$, i.e.
\begin{equation*}
S[U\phi]=S[\phi],\quad U\in G\,,
\end{equation*}
then every critical point of $S$, restricted to the subset of 
those $\phi$ which do not change
under the group action, is automatically a critical point of $S$ 
on the space of all fields. Now we choose for $G$ the little group of 
$e_1$ and this proves that a critical point on $\cH_1$ automatically
solves the field equation in the full space, i.e. is a bounce
solution (an ansatz $\phi\parallel e_1$ is consistent).

Now we use the (in)equalities \eqref{B.2} and \eqref{B.3} to show that 
to every critical point on $\cH$ with finite action there exists a critical point on $\cH_1$  which has a smaller action.
So let us assume that $\phi\in\cH$ is a bounce solution 
of $S$. The virial theorem
\begin{equation}
S[\phi]=\frac{2T[\phi]}{d}\bigg\{\frac{2T[\phi]}{-d_c V[\phi]}
\bigg\}^{d/d_c},\qquad d_c=\frac{2d}{d-1}\,,
\label{B.4}
\end{equation}
together with \eqref{B.2} and \eqref{B.3} implies
\begin{equation}
S[\phi]\geq \frac{2T[\chi]}{d}\bigg\{\frac{2T[\chi]}{-d_c V[\chi]}
\bigg\}^{d/d_c}\,.\label{B.5}
\end{equation}

Next we use the result (see \cite{14}) that $S$, restricted to $\cH_1$, 
has a \emph{spherically symmetric and monotonically decreasing} 
tunnel solution 
$\eta$ and $S[\eta]$ is given by the variational principle
\begin{equation}
S[\eta]=\inf_{\zeta\in\cH_1\atop V[\zeta]<0}
\frac{2T[\zeta]}{d}\bigg\{\frac{2T[\zeta]}{-d_c V[\zeta]}
\bigg\}^{d/d_c}\,.
\label{B.6}
\end{equation}

Because of the virial theorem the potential energy of a bounce
$\phi$ must be negative. With 
\eqref{B.2} the function $\chi$ is admissible on the right-hand side in 
\eqref{B.6}. Together with \eqref{B.5} we conclude
\begin{equation*}
S[\phi]\geq S[\eta]\,.
\end{equation*}
In going through the inequalities one sees that the equality sign holds exactly 
if $\phi$ is a spherically symmetric and monotonically decreasing one-component
field.

\end{document}